\documentclass[11pt]{JHEP3}
\usepackage{epsfig}
\usepackage{axodraw}
\def\be{\begin{equation}}
\def\ee{\end{equation}}
\def\bea{\begin{array}}
\def\eea{\end{array}}
\def\beqa{\begin{eqnarray}}
\def\eeqa{\end{eqnarray}}
\def\beqas{\begin{eqnarray*}}
\def\eeqas{\end{eqnarray*}}

\def\bp{\begin{picture}}
\def\ep{\end{picture}}
\def\bc{\begin{center}}
\def\ec{\end{center}}
\def\bfig{\begin{figure}}
\def\efig{\end{figure}}

\def\bit{\begin{itemize}}
\def\eit{\end{itemize}}
\def\nn{\nonumber}
\def\f{\frac}

\def\dis{\displaystyle}
\def\[{\left[}
\def\]{\right]}
\def\({\left(}
\def\){\right)}

\def\..{\left.}
\def\.{\right.}

\def\la{\leftarrow}

\def\dif{\partial}

\def\da{\dagger}
\def\LL{{\cal L}}
\def\DD{{\cal D}}
\def\cut{\Lambda}
\def\la{\lambda}

\def\al{\alpha}
\def\bt{\beta}
\def\ka{\kappa}
\def\ze{\zeta}
\def\si{\sigma}
\def\ep{\epsilon}
\def\rh{\rho}
\def\et{\eta}
\def\de{\delta}

\def\pa{\partial}

\title{Vacuum Stability Bounds On Higgs Mass With Gravitational Contributions}
\author{Fei Wang$^{1,2}$\\
$^1$International Joint Research Laboratory for Quantum Functional Materials of Henan Province,
and Department of Physics and Engineering, Zhengzhou University, Henan 450001, P. R. China\\
$^2$  Kavli Institute of Theoretical Physics China, Chinese Academy of Sciences, Beijing 100190, P. R. China
}

\abstract{  We calculate the gravitational contributions to $\phi^4$ theory with general $R_\xi$ gauge-fixing choice and find that the result is gauge independent. Based on weak coupling expansion of gravity and ignoring the possible higher dimensional operators from "integrating out" the impact of gravity, we study the impacts of gravitational effects on vacuum stability.
New contributions to the beta function of scalar quartic coupling $\la$  by gravitational effects can modify the RGE running of $\la$ near the Planck scale. Numerical calculations show that the lower bound of higgs mass requiring absolutely vacuum stability can be relaxed for almost 0.6 to 0.8 GeV depending on the choice of top quark mass.
}

\begin{document}
\maketitle \indent
\newpage
\section{Introduction}

   We know that both the ATLAS and CMS collaborations of the Large Hadron Collider(LHC) experiment have established the existence of a 125 GeV Standard Model-like higgs boson\cite{atlas,cms}. The LHC data on higgs boson(with large uncertainties) agree well with the Standard Model predictions (except the possible enhanced diphoton signal by ATLAS) and no signs of new physics beyond the standard model are observed so far. However, naively extending the validity range of the standard model from electroweak scale to Planck scale maybe problematic. In addition to the aesthetic problem related to quadratic divergence scalar mass, the renormalization group equation (RGE) running of quartic coupling $\la$ with current higgs mass data of LHC changes into negative values near Planck scale which will lead to vacuum instability. In fact, as indicated in \cite{higgsbound,higgsbound2},
absolute stability of the higgs potential is excluded at 98\% C.L. for $M_h < 126 {\rm GeV}$. Similar result is obtained in \cite{lindner}.

      Negative $\la$ could lead to another local minimum at large field value.  If the new minimum lies below the electroweak (EW) vacuum, quantum tunneling effects from EW vacuum to the deeper one could make vacuum decay. It is in principle possible for us to live in a universe with metastable vacuum if the lifetime of such local minimum is larger than the age of our universe. In fact, current central value of higgs mass $m_H=125.9$ GeV given by LHC can lead to a metastable vacuum with long-enough lifetime\cite{metastablebound}. Even though the metastable scenario could be phenomenological acceptable, such scenario is not satisfying and there still exist the possibility of cosmic ray collision induced fast vacuum decay\cite{sher,lindner2,sher2}. So absolute vacuum stability is still the most appealing scenario for theoretical physicists.

     In order to reconcile the observed (low) higgs mass with absolute vacuum stability requirement, one can change the UV behavior of quartic couplings by introducing many well motivated new physics models beyond the Standard Model\cite{NP1,NP2}. However, an important ingredient in Standard Model which had not received enough attention is gravity. Although gravitational effects decouple in most of the discussions related to standard model, such effects can be important near the Planck scale which may change the RGE running behavior of quartic coupling in the UV region. An interesting consequence of gravitational effects is the asymptotic free behavior of all gauge couplings near Planck scale when new power-law running gravitational contributions become dominant\cite{wilczek}.

     Authors in \cite{gaugedependent1,gaugedependent2} found that the calculation by \cite{wilczek} with background field method are in general gauge dependent and the true contribution vanishes. Further studies\cite{ylwu,tom,hongjian-he,kiritsis,daum} again confirms the non-zero effects for the running of gauge couplings by \cite{wilczek}.  The gravitational contribution to scalar and yukawa theory is also calculated with Vilkovisky-DeWitt method\cite{hongjian-he,tom2} or various methods with specified gauge fixing condition\cite{scalar0,scalar1,scalar2,scalar3,artur,sunyi}. However, it is important to check the gauge dependence of gravitational contributions in order to get physical results.
We carry out the calculation with traditional Feynman diagram methods and check that our result is gauge independent.  With the gauge independent results on quartic coupling beta functions, we could study the gravitational new contributions on vacuum stability problem. Endeavors along this line can be found in\cite{vacuumGR} in which the higgs mass was "predicted" even before the higgs was discovered. We use a different approach and discuss the status of higgs mass lower bound (from absolute vacuum stability requirement) with new gravitational contributions.

     This paper is organized as follows. In sec 2, we perform the calculation of gravitational contributions to quartic coupling with the most general gauge-fixing  choice. In sec-3, we discuss the effects of such gravitational contributions to vacuum stability problem. Sec-4 contains our conclusion.

 \section{Gravitional Corrections to Scalar $\Phi^4$ Coupling}
 It is well known that quantum gravity is nonrenormalizable. However, as pointed out in \cite{donoghue} on general relativity as an effective theory, physical predictions for such a nonrenormalizable theory is justified if we are only interest in physics at a scale $E\ll M_{Pl}$. The predictions should coincide with the results given by the underlying fundamental theory whatever its nature. So the resulting power law running of $\lambda$, which will be given shortly, should be interpreted to hold in the validity range of such an effective theory.

   The action $S$ for the scalar-gravity system can be written as,
 \begin{eqnarray}
 S &=&\int d^dx\sqrt{-g}\[\kappa^{-2}R
 -\f{1}{2\ze}\(\dif_\nu h^{\mu\nu}-\f{1}{2}\dif^\mu h\)^2
 + \LL_\phi +\cdots \]
 \\
 \LL_\phi &=&
 \frac{1}{2}g^{\mu\nu}\dif_{\mu}\phi\dif_{\nu}{\phi}
 -\frac{1}{2}m^2\phi^2
 -\frac{1}{4!}{\lambda}\phi^4
 \end{eqnarray}
 where $R$ is the Ricci scalar curvature. The gravitational coupling
 $\,\ka =\sqrt{16\pi G_N}\approx (1.69 \times 10^{18} {\rm GeV})^{-1}$
are determined by the Newton constant $G_N$ with the Planck scale given by
  $\,{M}_{\rm Pl}=G_N^{-\f{1}{2}}
   \simeq 1.22\times 10^{19}\,$GeV.\,

 We make the weak-coupling expansion for the Einstein gravity,
 \beqa
 g_{\mu\nu} &=& \eta_{\mu\nu} +\ka h_{\mu\nu} \,,
 \nn\\
 g^{\mu\nu} &=& \eta^{\mu\nu} -\ka h^{\mu\nu}
                 +\ka^2 h^{\mu\si}h^{\nu}_{\si} +O(\ka^3) \,,
 \\
 \sqrt{-g} &=& 1+\f{\ka}{2}h
                +\f{\ka^2}{8}(h^2-2h^{\mu\nu}h_{\mu\nu})
                +O(\ka^3),
 \nn
 \eeqa
 where
 \,$\eta_{\mu\nu}=\eta^{\mu\nu}=(1,-1,-1,-1)$\, and
 \,$h=h^{\mu\nu}\eta_{\mu\nu}=h^\mu_\mu$\,.\,

 Now let us expand the action up to $O(\ka^2)$,
 \beqa
 \label{eq:L-kappa}
 S & \equiv & \int d^dx \LL
 ~=\int d^dx \[
 \LL_{(0)} + \LL_{(1)} + \LL_{(2)} + O(\ka^3) \]
 \\
 \label{eq:L-kappa0}
 \LL_{(0)} &=&-\f{1}{2}\pa_\la
h^{\la\mu}\pa_\mu h^\nu_\nu+\f{1}{2}\pa_\la h^{\la\mu}\pa^\nu
h_{\mu\nu}-\f{1}{4}\pa_\la h^{\mu\nu}\pa^\la
h_{\mu\nu}+\f{1}{4}\pa_\la h^\mu_\mu\pa^\la h^\nu_\nu
 \nn\\
 &&
 -\f{1}{2\zeta}\(\pa_\nu h^{\mu\nu}-\f{1}{2}\pa^\mu h \)^2+\f{1}{2}\(\dif_\mu\phi\dif^\mu\phi -m^2\phi^2\)
 -\f{\la}{4!}\phi^4
 \\
 \label{eq:L-kappa1}
 \LL_{(1)} &=&
 \f{\ka}{2}\(\f{1}{2}\et^{\mu\nu}h-h^{\mu\nu}\)
  \dif_\mu\phi\dif_\nu\phi
  -\f{\,\ka m^2}{4}h\phi^2
  -\f{\ka\la}{2\cdot 4!}h\phi^4 + O(\ka h^3)
 \\
 \label{eq:L-kappa2}
 \LL_{(2)} &=&
 \f{\ka^2}{2}
 \[\f{1}{8}\(h^2-2h^{\si\rh}h_{\si\rh}\)\et^{\mu\nu}
   -\f{1}{2}hh^{\mu\nu}
   +h^{\mu\si}h^\nu_\si\]\dif_\mu\phi\dif_\nu\phi
 \nn\\
 &&
   -\f{\,\ka^2m^2}{16}\(h^2-2h^{\mu\nu}h_{\mu\nu}\)\phi^2
   -\f{\ka^2\la}{8\cdot 4!}
    \(h^2-2h^{\mu\nu}h_{\mu\nu}\)\phi^4
  +O(\ka^2h^4) \,.
 \eeqa
  We can thus derive the free graviton propagator,
 \beqa
 \DD_{\mu\nu,\si\rh}(p) &=& \f{i}{p^2+i\ep} \[
 \eta_{\mu\si}\eta_{\nu\rh} +
 \eta_{\mu\rh}\eta_{\nu\si} -
 \f{2}{d-2}\et_{\mu\nu}\et_{\si\rh} \right.
 \nn\\[3mm]
 &&\hspace*{15mm} \left.
 - \f{1-\ze}{p^2}\(
 p_\mu p_\si \et_{\nu\rh} +
 p_\mu p_\rh \et_{\nu\si} +
 p_\nu p_\si \et_{\mu\rh} +
 p_\nu p_\rh \et_{\mu\si}
 \)
 \]
 \eeqa
 where $d=4$ is the space-time dimension.

  The relevant one-loop self-energy and vertex diagrams are
 shown in Fig.\,1-2. It is important to note that only one graviton
 propagator appears in each diagram where all the
 vertex-couplings are independent of the gauge-fixing
 parameter $\ze$.

 \begin{figure}[htbp]
\includegraphics[width=4.0in,height=1.0 in]{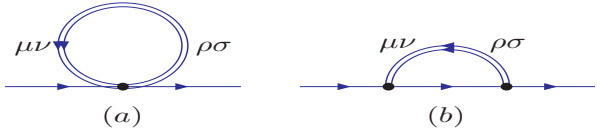}
\label{fig22}
 \caption{One-loop graviton corrections to the scalar
 self-energy with quadratic divergence.}
 \end{figure}

 We first compute the scalar self-energy in Fig.\,1(a). The relevant
 Feynman rules can be seen in appendix {\bf A}.
 \beqa
 i\Pi(q^2)[a] &=&
\f{1}{2} \int_p C_4^{\mu\nu,\si\rh}[\phi(q)\phi(-q) hh]
 \DD_{\mu\nu,\si\rh}(p)
 \nn\\
 &=& -q^2\f{\ka^2}{4}(d-4)\[(d-1)+2\ze\]iI_2\,
 \label{eq:SF-a}
 \eeqa
 where $\,\dis\int_p \equiv \int \f{d^dp}{(2\pi)^d}\,$ and
 the loop-integral $I_2$ is defined as
 \beqa
 \label{eq:I2}
 iI_2 &\equiv& \int \f{d^dp}{(2\pi)^d} \f{1}{\,p^2\,}
 \eeqa
 which is quadratically divergent at $d=4$.  The whole contribution
 of Fig.\,1(a) to the self-energy (\ref{eq:SF-a}) vanishes identically at $d=4$.
 The above coupling $C_4^{\mu\nu,\si\rh}[\phi(q)\phi(-q) hh]$
 for the $\phi-\phi-h-h$ vertex is derived in the appendix.
 Next we compute the scalar self-energy in Fig.\,1(b),
 \beqa
 i\Pi(q^2)[b] &=&
 \int_p
 C_3^{\mu\nu}[\phi(q)\phi(-p-q)h]
 C_3^{\si\rh}[\phi(p+q)\phi(-q)h]
 \DD_{\mu\nu,\si\rh}(p)
 \nn\\
 &=&{q^2}\(\ze \ka^2 iI_2\) \,.
 \label{eq:SF-b}
 \eeqa
 In summary, we deduce the following total scalar
 self-energy contributions (with quadratic divergence),
 \beqa
 i\Pi(q^2) &=& i\Pi(q^2)[a] + i\Pi(q^2)[b]
 \nn\\
 &=& q^2\f{\ka^2}{8}\[(d-4)(d-1)+2d\ze\]iI_2 \nn\\
 &=& q^2\ka^2 \zeta iI_2\,
 \eeqa
 where only the $\ze$-dependent term survives at $d=4$.
The scalar wavefunction renormalization
 \,$\phi_0 = Z_\phi^{\f{1}{2}}\phi$\,
 provides the counter term for the self-energy
 renormalization,
 \beqa
 \Pi_r(q^2) &=& \Pi(q^2) + q^2(Z_\phi -1) \,,
 \nn\\
 \de Z_\phi{\equiv}Z_\phi -1
 &=& \f{\ka^2}{8}\[(d-4)(d-1)+2d\ze\]
 \[I_2(E)-I_2(\cut) \]\nn\\
 &=& \ka^2 \ze \[I_2(E)-I_2(\cut) \].
 \eeqa
 where $E$ is the renormalization scale and
 \beqa
 I_2(\cut ) &=& -i\int_0^\cut\f{d^dp}{(2\pi)^d}\,
 \f{1}{p^2} \,=\, -\f{\cut^2}{\,16\pi^2\,} \,,
 ~~~~({\rm for}~d=4)\,.
 \eeqa

 We will not specify how this integral is regularized until we explicitly prove the $\ze$-cancellation for the exact gauge-invariance.
The proof of gauge-invariance ($\ze$-cancellation) of the one-loop $\bt$-function does not depend on the explicit form of the integral $I_2$ (except that we are sure
 that it can be properly regularized).
 \begin{figure}[htbp]
\includegraphics[width=4.0in,height=1.0 in]{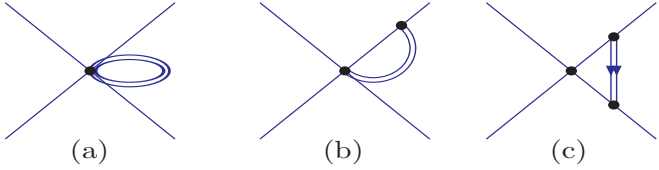}
\label{fig22}
\caption{One-loop graviton corrections to the 4-scalar
  vertex with quadratic divergence. The first diagram contributes while the last two diagrams give null contributions.}
\end{figure}
 Finally, we compute the graviton-induced correction to the scalar-vertex $\la\phi^4$, which is given by the loop diagram in Fig.\,2. We checked that the last two diagrams in Fig.\,2 give null contributions. Our calculation gives

 \beqa
 i\Gamma_4 &=& i\f{1}{2}\f{\ka^2\la}{4}
 \(\et^{\mu\si}\et^{\nu\rh}+
   \et^{\mu\rh}\et^{\nu\si}-
   \et^{\mu\nu}\et^{\si\rh}\)
   \int_p \DD_{\mu\nu,\si\rh}(p)
 \nn\\[2mm]
 &=&
 -\f{\ka^2\la}{4}(d^2-d+2d\ze )iI_2(\cut) \,,\nn\\
 &=&-{\ka^2\la}(3+2\ze) i I_2\,
 \eeqa
 where the counter term for vertex renormalization
 can be derived from renormalization condition
 $\delta_\la=\la_0 Z_\phi^2-\la$ and Feynman rules for counter
 terms:
\beqa i\Gamma_{4r}=i\Gamma_4-i \de_\la \eeqa
 \beqa
\de_\la &=& \f{\ka^2\la}{4}(d^2-d+2d\ze )
 \(I_2(E)-I_2(\cut)\) \nn\\
 &=&\ka^2\la(3+2\ze)\(I_2(E)-I_2(\cut)\).
 \eeqa

  We have checked that the mass terms does not change the previous
expression for quadratic divergence terms.
 From the wavefunction and vertex renormalizations in the
 previous section, we can compute the
 renormalization for the quartic scalar coupling $\la$ at
 one-loop,
 \beqa
 \la &=& \la_0  Z_\phi^{2}-\de_\la
\nn\\[1mm]
 &=& \la_0 \left\{1+{\ka^2}\[1-d\]
  \[I_2(E)-I_2(\cut)\]\right\}
 \\[2mm]
 &=& \la_0 \[1 +\f{3\ka^2}{16\pi^2}\(E^2-\cut^2\)\] \,,
 ~~~~~({\rm for}~d=4)\,.
 \eeqa

So we get the gravitational contributions to scalar beta functions
   \beqa
 \,\bt(\la)  &=& \la \f{\dif \ln\la}{\dif\ln E}
 \,=\, \f{3\la\kappa^2}{8\pi^2}E^2\,,\,\eeqa

 In a theory with complex scalar fields,the Lagrangian reads:
\begin{eqnarray*}
{\cal L}=\int d^4x\sqrt{-g}(\kappa^{-2}
R+\partial_{\mu}\phi^{\dagger}\partial_{\nu}{\phi}g^{\mu\nu}-m^2\phi^{\dagger}\phi+{\lambda}(\phi^{\dagger}\phi)^2+\frac{1}{2\eta\kappa^2}[\partial_{\mu}(\sqrt{-g}g^{\mu\nu})]^2+\cdots)
\end{eqnarray*}
Define  $\phi=\frac{1}{\sqrt{2}} (\phi_3+i\phi_4, \phi_1+i\phi_2)$, then we have $\phi^2=\phi^{\dagger}\phi=\frac{1}{2}\sum\limits_{i=1}^4\phi_i^2$
with the Vacuum Expectation Value (VEV) $<\phi^2>_0=\frac{1}{2}<\phi_1>_0^2=\frac{1}{2}v^2$. The effective potential can depend only on $\phi^2$, it will be
adequate to calculate the loop diagrams with external $\phi_1$.
Expanding the expression according to $\phi_1$ gives
\beqa -\frac{1}{2}m^2\phi_1^2+\frac{\la_{EW}}{4}\phi_1^4+\cdots
\eeqa So the contribution of graviton differs from $\phi^4$ theory
with the replacement \beqa \la_{\phi^4}{\rightarrow}\frac{1}{6}\la_{EW} \eeqa
Goldstone bosons do not give contributions to gravitational loops. The total
gravitational contributions have the same form in both theories (complex and real scalar theories):
\beqa
 \label{eq:Beta}
 \bt(\la) &=& \la \f{\dif \ln\la}{\dif\ln E}
 \,=\, \beta_{\rm gauge} +\beta_{\rm scalar}+\f{3\la}{\,8\pi^2\,} \ka^2E^2,
 \,
\eeqa

 It is possible that a more general interaction Lagrangian for scalar-gravity system
can contain the renormalizable non-minimal coupling term (Veltman term):
 \beqa
\Delta S=\xi\int d^dx \sqrt{-g}\{ R \phi^2\}
 \eeqa
with $'R'$ the Ricci scalar curvature and $'\xi'$ an arbitrary dimensionless parameter. Such coupling is renormalizable and will always appear in the Lagrangian. For example, this term will always appear in the loop level\cite{xiterm}. The value of the dimensional parameter $\xi$ can be constrained by LHC data\cite{calmet} as well as the unitary bound of $W_LW_L$ scattering\cite{hj-unitary}.
So we should see if the presence of such terms can have new  effects on
the scalar $\beta$-functions. Careful analysis indicates that the presence of Veltman term can only contribute to the wave function renormalization.
However, detailed calculations shows that such contributions to wave function renormalization vanish.  Therefore, the gravity contributions to scalar beta function will not be changed with the presence of such non-minimal couplings.

\section{Vacuum Stability Bounds With Gravitational Contributions}
 The presence of a new term in the scalar quartic beta function from gravitational effects can have important consequences. Such term can be dominant near the Planck mass scale and significantly change the running behavior of quartic coupling in the UV region. Thus, it is possible that the vacuum stability problem can be ameliorated by taking into account this gravitational contributions. In fact, improvement of vacuum stability lower bounds on higgs mass of order $0.1~{\rm GeV}$ from additional gravitational contributions could be important because they can be comparable to certain higher loop improvements.

 As noted previously, the power law running of $\lambda$ will be valid in the sense of effective field theory with energy scale well below the Planck scale. On the other hand, there are indications that gravity is asymptotic safe. The effective Planck scale $M_{pl}^2(k)$ can be changed with respect to the characteristic energy scale $M_{pl}^2(k)=M_{Pl}^2+2\xi_0 k^2$. So it is possible that the high energy scattering amplitude involving the effective gravitational constant behave regularly for momentum transfer $k^2\gg M_{Pl}^2$.  At the same time, the power law running of the $\lambda$ should also be modified with the ratio $k^2/M_{pl}(k^2)$ in the beta function tending to a positive constant $1/2\xi_0$ when $k^2\gg M_{Pl}^2$. In fact, with positive gravity induced anomalous dimension $A_\la>0$ which is just our cases, $\la$ at scales beyond the Planck mass scale is determined by a fixed point at zero. Thus the physical higgs mass is predicted very close to the lower bound of the infrared interval for $\la$. Detailed discussions can be seen in \cite{vacuumGR}. So the blowing up behavior of $\la$ from gravitational contributions in our following conclusions will be changed upon the Planck mass scale. We only concentrate on the small gravitational contributions  to $\la$ in the regime with $k^2\lesssim M_{Pl}/\sqrt{2\xi_0}$. 

Besides, nonrenormalizable scalar couplings can also have important consequences on the vacuum stability problems. As noted in \cite{scalarNR1,scalarNR2}, the standard model scalar sector augmented by dimension-6 and dimension-8 scalar coupling could modify the stability condition of the electroweak vacuum. The presence of such higher dimensional operators could also change the power law running blowing-up behavior of $\la$ for energy scale one or two orders below the suppression scale. Because of uncertainties in the new physics interactions at the Planck scale, we neglect such higher dimensional operators and take into account only the contributions from weak coupling expansion of gravity. Even though such gravitational contributions are non-complete, they could be crucial to keep $\la$ positive up to scale one order below the Planck scale.

Our numerical calculations indicate that positivity requirement of $\lambda$ at Planck mass scale (in case without considering gravitational effects) changes into the requirement that $\lambda$ is positive at $0.5 \times 10^{17}{\rm GeV}$ in our scenario. In order to study the RGE running of quartic coupling $\la$, we adopt the full two-loop Standard Model beta functions\cite{twoloop} for $\la$(three loop results can be seen in\cite{threeloop1,threeloop2}), the top-yukawa couplings $y_t$ and gauge couplings $g_i (i=1,2,3)$ in the region between $m_{top}$ and $\ka^{-1}$ in addition to the one-loop power-law-running contribution terms from gravitational effects. The following boundary conditions\cite{pdg} for RGE running
\beqa
\al_s(M_Z)&=&0.1184\pm0.0007\nn\\
\al_{\rm em}^{-1}(M_Z)&=&127.906\pm0.019\nn\\
\sin^2\theta_W(M_Z)&=&0.2312\pm0.0002,\nn\\
M_{\rm higgs}&=&125.9\pm0.4{\rm  GeV}.\nn\\
y_b(M_Z)&=&0.0162834,\nn\\
y_\tau(M_t)&=&0.0102.
\eeqa
are used and we obtain
\beqa
\al_2(M_Z)&=&\al_{em}(M_Z)/\sin^2\theta_W=(29.5718)^{-1},\nn\\
\al_1(M_Z)&=&\al_{em}(M_Z)/\cos^2\theta_W=(98.3341)^{-1}.
\eeqa
The two loop RGE running for gauge couplings are given by
\beqa
\f{d }{d\ln E}g_i=\f{b_i}{16\pi^2}g_i^3+\f{g_i^3}{(16\pi^2)^2}\[\sum\limits_{k}b_{ki}g_k^2-Tr\(C_k^U F_U^\da F_U+C_k^D F_D^\da F_D+C_k^L F_L^\da F_L\)\].
\eeqa
with
\beqa
b_{ki}=\(\bea{ccc}\f{199}{50}&\f{9}{10}&\f{11}{10}\\\f{27}{10}&\f{35}{6}&\f{9}{2}\\\f{44}{5}&12&-26\eea\)
\eeqa
 and the yukawa matrix
\beqa
C_k^U=\(\f{17}{10},\f{3}{2},2\)~, C_k^D=\(\f{1}{2},\f{3}{2},2\)~,C_k^L=\(\f{3}{2},\f{1}{2},0\).
\eeqa
The normalization $g_1^2=\f{5}{3}g_Y^2$ is used in previous expressions. We keep only the yukawa coupling of the third generation and neglect the sub-leading contributions from the first two generations. So we can write explicitly the RGE running for $g_Y$
\beqa
\f{d g_Y}{d\ln E}=\f{1}{16\pi^2}\f{41}{6}g_Y^3+\f{1}{(16\pi^2)^2}\[\f{199}{18}g_Y^5+\f{9}{2}g_Y^3g_2^2+\f{44}{3}g_Y^3g_3^2-\f{17}{6}y_t^2g_Y^3-\f{5}{6}y_b^2g_Y^3-\f{5}{2}y_\tau^2g_Y^3\].\nn
\eeqa
and $g_2$
\beqa
\f{d g_2}{d\ln E}=-\f{1}{16\pi^2}\f{19}{6}g_2^3+\f{1}{(16\pi^2)^2}\[\f{3}{2}g_Y^2g_2^3+\f{35}{6}g_2^5+12g_2^3g_3^2-\f{3}{2}y_t^2g_2^3-\f{3}{2}y_b^2g_2^3-\f{1}{2}y_\tau^2g_2^3\].\nn
\eeqa
as well as $g_3$
\beqa
\f{d g_3}{d\ln E}=-\f{1}{16\pi^2}7g_3^3+\f{1}{(16\pi^2)^2}\[\f{11}{6}g_Y^2g_3^3+\f{9}{2}g_2^2g_3^3-26 g_3^5-2y_t^2g_3^3-2y_b^2g_3^3\].
\eeqa

We also include two-loop top-yukawa RGE
\beqa
\f{d}{d\ln E}F^U=\f{1}{16\pi^2}\beta_U^1F^U+\f{1}{(16\pi^2)^2}\beta_U^2F^U~,
\eeqa
with
\beqa
\beta_U^1=\f{3}{2}\( F_U^\da F^U-F_D^\da F_D\)+Tr(3F_U^\da F_U+3F_D^\da F_D+F_L^\da F_L)-\(\f{17}{20}g_1^2+\f{4}{9}g_2^2+8g_3^2\),
\eeqa
and a lengthy expression for $U^2$. The simplified expression reads
\beqa
\f{d}{d\ln E} y^t&=&\f{1}{16\pi^2}\(\f{9}{2}y_t^3+\f{3}{2}y_b^2y_t+y_{\tau}^2y_t-8g_3^2y_t-\f{4}{9}g_2^2y_t-\f{17}{12}g_Y^{ 2}y_t\)\nn\\&+&\f{1}{(16\pi^2)^2}\[-12y_t^4-\f{11}{4}y_t^2y_b^2-\f{1}{4}y_b^4+\f{5}{4}y_b^2y_\tau^2-\f{9}{4}y_t^2y_\tau^2
-\f{9}{4}y_\tau^4+6\la^2-12\la y_t^2-4\la y_b^2\.\nn\\
&+&(\f{403}{48}g_Y^2+\f{225}{16}g_2^2+36g_3^2)y_t^2
+\(\f{7}{48}g_Y^2+\f{99}{16}g_2^2+4g_3^2\)y_b^2+\(\f{25}{8}g_Y^2+\f{15}{8}g_2^2\)y_{\tau}^2\nn\\
&+&\left.\f{1187}{216}g_Y^4-\f{3}{4}g_Y^2g_2^2+\f{19}{9}g_Y^2g_3^2-\f{23}{4}g_2^4+9g_2^2g_3^2-108g_3^4\].
\eeqa
The contribution of bottom,tau yukawa couplings to $\la$ are negligible, so we use here only the one-loop results for bottom yukawa
\beqa
\f{d}{d\ln E}y^b=\f{1}{16\pi^2}\(\f{9}{2}y_b^3+\f{3}{2}y_t^2y_b+y_{\tau}^2y_b-\f{5}{12}g_Y^2y_b-\f{9}{4}g_2^2y_b-8g_3^2y_b\),
\eeqa
and tau-yukawa couplings
\beqa
\f{d}{d\ln E}y^{\tau}=\f{1}{16\pi^2}\(\f{5}{2}y_{\tau}^3+3y_t^2y_\tau+3y_b^2y_\tau-\f{15}{4}g_Y^2-\f{9}{4}g_2^2\).
\eeqa

The two-loop RGE for $\lambda$ are given by
\beqa
\f{d}{d\ln E} \lambda=\f{1}{16\pi^2}\beta_\la^1+\f{1}{(16\pi^2)^2}\beta_\la^2~,
\eeqa
 with
\beqa
\beta_\la^1&=&24\la^2-\(3g_Y^2+g_2^2\)\la+\f{9}{8}\(\f{1}{3}g_Y^4+\f{2}{3}g_Y^2g_2^2+g_2^4\)+4\la(3y_t^2+3y_b^2+y_\tau^2)\nn\\
 &-&2(3y_t^4+3y_b^4+y_\tau^4)+6\la \kappa^2 E^2~,
\eeqa
and
\beqa
\beta_\la^2&=&-312\la^3+36(g_Y^2+3g_2^2)\la^2-\(\f{73}{8}g_2^4+\f{39}{4}g_Y^2g_2^2+\f{373}{24}g_Y^4\)\la\nn\\
&+&\f{1}{2}\(\f{305}{8}g_2^6-\f{289}{24}g_2^4g_Y^2-\f{559}{24}g_2^2g_Y^4-\f{379}{24}g_Y^6\) \nn\\&-&32 g_3^2(y_t^4+y_b^4)-\f{4}{3}g_Y^2\(2y_t^4-y_b^4+3y_\tau^4\)-\f{3}{4}g_2^4(3y_t^2+3y_d^2+y_\tau^2)\nn\\
&+&10\la\[\(\f{17}{12}g_Y^2+\f{9}{4}g_2^2+8g_3^2\)y_t^2+\(\f{5}{12}g_Y^2+\f{9}{4}g_2^2+8g_3^2\)y_b^2+\(\f{5}{4}g_Y^2+\f{3}{4}g_2^2\)y_\tau^2\]\nn\\
&+&\f{g_Y^2}{2}\[\(21g_2^2-\f{19}{2}g_Y^2\)y_t^2+\(9g_2^2+\f{5}{2}g_Y^2\)y_b^2+\(11g_2^2-\f{25}{2}g_Y^2\)y_\tau^2\]-48\la^2(3y_t^2+3y_b^2+y_\tau^2)\nn\\
&-&\la(3y_t^4+3y_b^4+y_\tau^4)+6\la y_t^2y_b^2+10\(3y_t^6+3y_b^6+y_\tau^6\)-6y_t^4y_b^2-6y_t^2y_b^4.
\eeqa
with the one-loop RGE equation modified by gravitational contributions.

 We adopt the initial value obtained in \cite{higgsbound} with $\overline{\rm MS}$ scheme renormalized at the pole top mass
\beqa
\la(m_t)&=&0.12577 + 0.00205\(\f{m_{higgs}}{\rm GeV}-125\)-0.00004\(\f{m_t}{\rm GeV}-173.15\)\pm 0.00140_{th},\nn\\
y_t(m_t)&=&0.93587 + 0.00557\(\f{m_t}{\rm GeV}-173.15\)-0.00003\(\f{m_{higgs}}{\rm GeV}-125\)\nn\\
      &-&0.00041\(\f{\al_s(M_Z)-0.1184}{0.0007}\)+0.00200_{th},\nn\\
g_3(m_t)&=& 1.1645 + 0.0031\(\f{\al_s(M_Z)-0.1184}{0.0007}\)-0.00046\(\f{m_t}{\rm GeV}-173.15\).
\eeqa

Typical benchmark points of $(m_{top}, m_{higgs}, \al_s(M_Z))$ values to illustrate the importance of gravitational effects are shown in fig.3. We can see that the UV behavior of quartic coupling can indeed be greatly modified near the Planck scale. In fact, absolute vacuum stability will no longer need the the quartic coupling $\la$ to be positive at Planck mass scale $(M_{Pl}=1.22\times 10^{19} {\rm GeV})$. Instead, if $\la$ is positive at scale of order $10^{17}{\rm GeV}$,
it will in general not tend to negative when it continues running to $\ka^{-1}$ at which the effective theory of gravity breaks down
and UV completes to a typical full consistent quantum-gravity theory, for example, the asymptotically safe theory. The vacuum stability problem thus can be ameliorated by slightly relaxing the lower bound of the higgs mass.
\begin{figure}[htbp]
\includegraphics[width=6.0in,height=3.0 in]{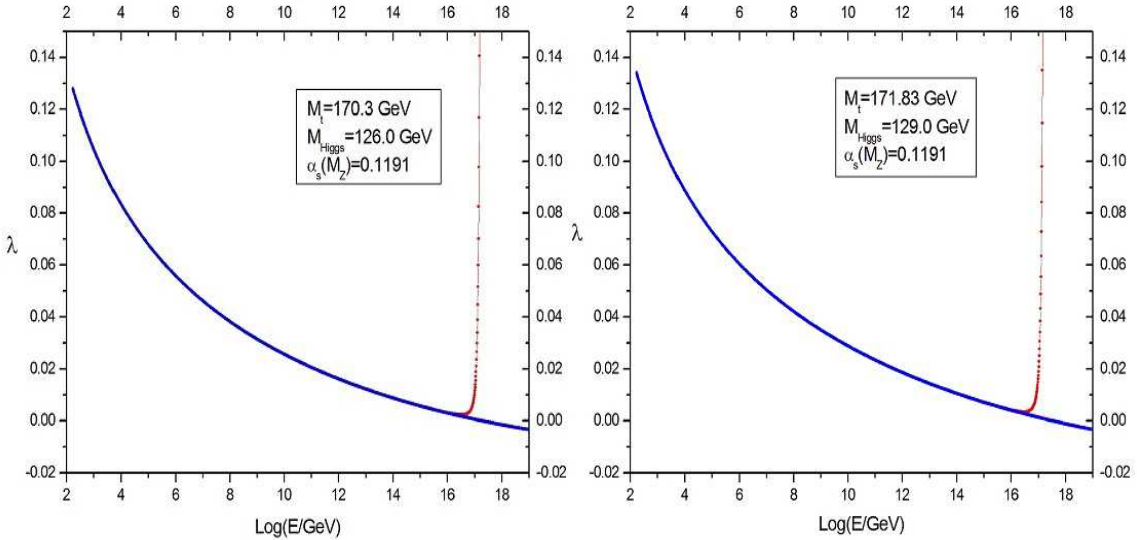}
\vspace{-0.5cm} \caption{The full two-loop RGE running of quartic coupling with (denoted by the red curve) and without (denoted by the blue curve) the gravitational contributions. The two curves are almost coincide with each other below the scale $E\approx 5.0\times 10^{16}$ GeV which indicates that the gravitational contributions become important above such scale.}
 \label{fig3}
\end{figure}

Unfortunately, our numerical results also show that current collider data on top and higgs mass can not be compatible with the absolute vacuum stability requirement even if the gravitational effects are taken into account. Various input shows that the gravitational contributions can relax the lower bound of higgs masses by 0.6-0.8 GeV. Improvements on lower bound of higgs boson for various choice of top quark mass can be found in table \ref{improvement}.

\begin{table}[htbp]
\caption{We fix $\al_s(M_Z)=0.1194$ and show the value of improvement (on lower bound of higgs mass) versus the top quark mass. Here $L_{higgs}^{Gr}$ ($L^{NGr}_{higgs}$) indicates the lower bound of higgs mass with(without) gravitational contributions.}
\begin{center}
\begin{tabular}{|c|c|c|c|c|c|}
\hline
 $M_{h} \slash  M_{t}$&174.31{\rm GeV}&173.07{\rm GeV}& 171.83{\rm GeV} &171.0{\rm GeV}&170.0{\rm GeV} \\
\hline $L_{higgs}^{Gr}$&133.6{\rm GeV} &131.2{\rm GeV}&128.75{\rm GeV}& 127.15{\rm GeV}&125.3{\rm GeV} \\
\hline $L_{higgs}^{NGr}$& 134.4{\rm GeV}& 131.9{\rm GeV}&129.45{\rm GeV}&127.75{\rm GeV}&125.9{\rm GeV}\\
\hline $\Delta m_h$&0.8 {\rm GeV}&0.7 {\rm GeV}& 0.7 {\rm GeV}&0.6{\rm GeV} &0.6 {\rm GeV}\\
\hline
\end{tabular}
\end{center}
\label{improvement}
\end{table}
\section{\label{sec-3}Conclusions}
   We calculate the gravitational contributions to $\phi^4$ theory with general $R_\xi$ gauge-fixing choice and find that the result is gauge independent.
Based on weak coupling expansion of gravity and ignoring the possible higher dimensional operators from "integrating out" the impact of gravity, we study the impacts of gravitational effects on vacuum stability. The beta function for quartic coupling $\la$  by gravitational effects can modify the RGE running of $\la$ near the Planck scale. Numerical calculations show that the lower bound of higgs mass requiring absolutely vacuum stability can be relaxed for almost 0.6 to 0.8 GeV depending on the choice of top quark mass.

 We should note again that the weak coupling expansion of gravity used in this paper is not sufficient. As noted before, contributions from the non-renormalizable aspects of gravity could be very important. The inclusion of certain higher-dimensional operators in higgs sector could modify significantly the behavior of the potential near the Planck scale and possibly destabilize the electroweak vacuum \cite{scalarNR1,scalarNR2}.  Furthermore, as noted previously, gravitational corrections to the suppression scale $M_{Pl}$ could also change the UV behavior of $\la$ near (and upon) the Planck scale if we assume asymptotic safe. Besides, change of the suppression scale in the higher dimensional operators from gravitational contributions could also cause certain effects.

\begin{acknowledgments}
Fei Wang acknowledges the referee for very helpful discussions. Fei Wang also acknowledges Prof. Hong-jian He for early stage cooperations on this work and very useful comments. This research was supported by the Natural Science Foundation of China under project 11105124,11105125.
\end{acknowledgments}

\section*{Appendix: Feynman rules}
We define the expression: \beqa
 C_{\mu\nu,\rho\sigma}=\et_{\mu\rho}\et_{\nu\sigma}+\et_{\mu\sigma}\et_{\nu\rho}-\et_{\mu\nu}\et_{\rho\sigma}
\eeqa
$\phi-\phi-h_{\mu\nu}$  vertex:
\begin{figure}[htbp]
\includegraphics[width=4.0in,height=1.0 in]{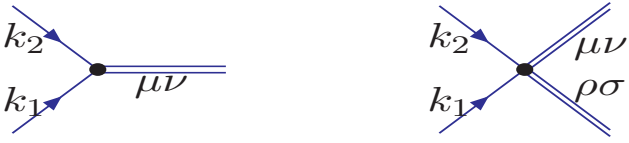}
\label{fig3}
\caption{Feynman rules for vertex $C_3^{\mu\nu}$(left) and  $C_4^{\mu\nu,\rho\sigma}$(right), respectively.}
\end{figure}
\begin{eqnarray}
C_3^{\mu\nu}[\phi(k_1)\phi(k_2)h]=-\frac{i
\kappa}{2}\left\{\eta_{\mu\nu}(k_1{\cdot}k_2)-(k_{1\mu}k_{2\nu}+k_{1\nu}k_{2\mu})
\right\}
\end{eqnarray}
With mass term, the Feynman rule can be written as:
\begin{eqnarray}
C_3^{\mu\nu}[\phi(k_1)\phi(k_2)h]=-\frac{i
\kappa}{2}\left\{\eta_{\mu\nu}(k_1{\cdot}k_2-m^2)-(k_{1\mu}k_{2\nu}+k_{1\nu}k_{2\mu})
\right\}
\end{eqnarray}

$\phi-\phi-h_{\mu\nu}-h_{\rho\sigma}$  vertex:
\begin{eqnarray*}
&&C_4^{\mu\nu,\rho\si}[\phi(k_1)\phi(k_2)hh]=i\frac{\kappa^2}{4}\left\{C_{\mu\nu,\rho\sigma}(k_1{\cdot}k_2)
+\eta_{\mu\nu}(k_{1\rho}k_{2\sigma}+k_{1\sigma}k_{2\rho})+\eta_{\rho\sigma}(k_{1\mu}k_{2\nu}+k_{1\nu}k_{2\mu})\right.\\
&&-\eta_{\mu\rho}(k_{1\nu}k_{2\sigma}+k_{1\sigma}k_{2\nu})
\left.-\eta_{\mu\sigma}(k_{1\nu}k_{2\rho}+k_{1\rho}k_{2\nu})-\eta_{\nu\sigma
}(k_{1\mu}k_{2\rho}+k_{1\rho}k_{2\mu})-\eta_{\nu\rho}(k_{1\mu}k_{2\sigma}+k_{1\sigma}k_{2\mu})\right\}
\end{eqnarray*}
with mass term:
\begin{eqnarray*}
&&C_4^{\mu\nu,\rho\si}[\phi(k_1)\phi(k_2)hh]=i\frac{\kappa^2}{4}\left\{C_{\mu\nu,\rho\sigma}(k_1{\cdot}k_2-m^2)
+\eta_{\mu\nu}(k_{1\rho}k_{2\sigma}+k_{1\sigma}k_{2\rho})+\eta_{\rho\sigma}(k_{1\mu}k_{2\nu}+k_{1\nu}k_{2\mu})\right.\\
&&-\eta_{\mu\rho}(k_{1\nu}k_{2\sigma}+k_{1\sigma}k_{2\nu})
\left.-\eta_{\mu\sigma}(k_{1\nu}k_{2\rho}+k_{1\rho}k_{2\nu})-\eta_{\nu\sigma
}(k_{1\mu}k_{2\rho}+k_{1\rho}k_{2\mu})-\eta_{\nu\rho}(k_{1\mu}k_{2\sigma}+k_{1\sigma}k_{2\mu})\right\}
\end{eqnarray*}
  When the term $a R \phi^2$ is included, the additional term for $\phi-\phi-h_{\mu\nu}-h_{\rho\sigma}$  vertex:
\beqas \delta C_4^{\mu\nu,\rho\si}[\phi(k_1)\phi(k_2)h(p_1)^{\mu\nu}h(p_2)^{\rho\sigma}]
&=&-4 a i\ka^2 \left\{-\f{1}{8}[(p_1^\rh p_2^\sigma+p_1^\sigma p_2^{\rh})\et^{\mu\nu}+(p_1^\mu
p_2^\nu+p_1^\nu p_2^\mu) \et^{\rho\sigma}]\.\\
&+&\f{1}{8}(p_1^\nu p_2^\rh\et^{\si\mu}+p_1^\mu
p_2^\rh\et^{\si\nu}+p_1^\nu
p_2^\si\et^{\rh\mu}+p_1^\mu p_2^\si\et^{\rh\nu})\\
&-&\f{1}{8}(p_1\cdot p_2)(\et^{\rh\nu}\et^{\si\mu}+\et^{\rh\mu}\et^{\si\nu})+\left.\f{1}{4}(p_1\cdot
p_2)\et^{\mu\nu}\et^{\rh\si}\right\}\eeqas

\begin{figure}[htbp]
\includegraphics[width=3.5 in,height=1.0 in]{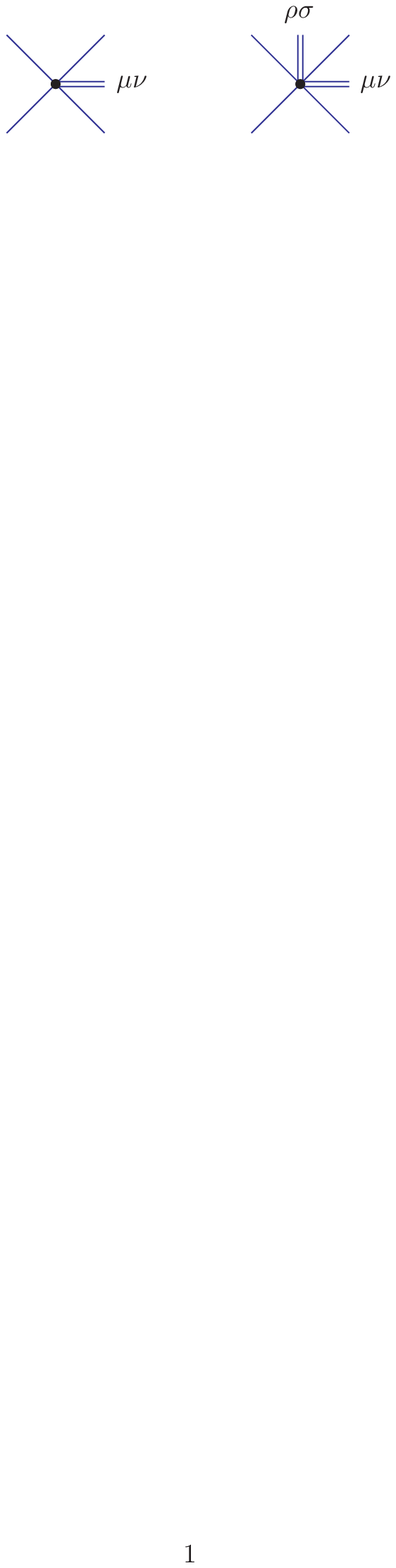}
\label{fig4}
\caption{Feynman rules for vertex $V_5^{\mu\nu}[\phi\phi\phi\phi h]$(left) and $V_6^{\mu\nu,\rho\sigma}[\phi\phi\phi\phi h h]$(right), respectively.}
\end{figure}

$\phi-\phi-\phi-\phi-h_{\mu\nu}$  vertex:
\begin{eqnarray}
V_5^{\mu\nu}[\phi\phi\phi\phi h]=\frac{i\kappa}{2}\lambda
\eta_{\mu\nu}\end{eqnarray}
$\phi-\phi-\phi-\phi-h_{\mu\nu}-h_{\rho\sigma}$  vertex:
\begin{eqnarray}
V_6^{\mu\nu,\rho\sigma}[\phi\phi\phi\phi h h]=\frac{i
\kappa^2}{4}\lambda C_{\mu\nu,\rho\sigma}
\end{eqnarray}

\end{document}